\documentclass[epj,twocolumn]{webofc}
\usepackage[varg]{txfonts}   
\woctitle{The innermost regions of relativistic jets and their magnetic fields}
\begin{document}
\title{On the connection between radio and gamma rays}
%
%
\subtitle{Variability and polarization properties in relativistic jets}

\author{M. Orienti\inst{1}\fnsep\thanks{\email{orienti@ira.inaf.it}} \and
        F. D'Ammando\inst{1} \and
        M. Giroletti\inst{1} \and
        D. Dallacasa\inst{2,1} \and
        T. Venturi\inst{1} \and 
        G. Giovannini\inst{2,1}
}

\institute{INAF-IRA, via Gobetti 101, 40129 Bologna, Italy
\and
           DIFA, University of Bologna, via Ranzani 1, 40127 Bologna, Italy
          }

\abstract{%
Relativistic jets are one of the most powerful manifestations of the
release of energy produced around supermassive black holes at the
centre of active galactic nuclei (AGN). Their emission is observed
across the entire electromagnetic spectrum, from the radio band to
gamma rays. Despite decades of efforts, many aspects of the physics of
relativistic jets remain elusive. In particular, the location and the
mechanisms responsible for the high-energy emission and the connection
of the variability at different wavelengths are among the greatest
challenges in the study of AGN.  

From the comparison of the radio and gamma-ray light curves of
gamma-ray flaring objects, there is evidence that some flares, 
either in radio or in gamma rays, have not an obvious connection at
the other extreme of the electromagnetic spectrum, like in the case of
the Narrow-Line Seyfert 1 SBS\,0846$+$513. An intriguing aspect pointed
out by high resolution radio observations is the change of the
polarization properties close in time with some high energy flares. In
particular, in PKS\,1510$-$089 and 3C\,454.3 a rotation of almost 90
degrees has been observed after strong gamma-ray flares. The swing of
the polarization angle may be related either to the propagation of a
shock along the jet that orders the magnetic field, or a change of the
opacity regime. 
}
\maketitle
\section{Introduction}
\label{intro}

The extragalactic gamma-ray sky is largely dominated by radio-loud
active galactic nuclei (AGN). In particular, the population of blazars
represents almost 97\% of the gamma-ray emitting AGN from the 2LAC
catalogue \citep{ackermann11}. The high energy
emission is likely due to inverse Compton scattering off the low energy
photons by the relativistic
electrons which are also responsible 
for the synchrotron emission observed in the
radio band. Although in blazars all the ingredients needed for this
scenario are present, a clear connection between the emission at
the edges of the 
multiwavelength spectrum has not been unambiguously established yet. In
particular, the trigger of the high-energy
flares typically observed in blazars, and the location of the
gamma-ray emitting region are still uncertain.\\
Thanks to their luminosity variability observed throughout the electromagnetic
spectrum, we have a chance to shed a light on this issue. The
variability behaviour shown in the various energy bands
(i.e. time-delay, duration, intensity) provides us tight constraints
on the location and size of the gamma-ray emitting region. For example,
intra-day variability is an indication of a very compact region.
Moreover, a time delay 
in the emergence of the flare at progressively longer wavelengths
may locate the high-energy emitting region in 
the innermost part of the AGN where
severe opacity effects play a role \citep{ghisellini96}. On the other
hand, the detection of gamma-ray and millimeter-wavelength 
flares occurring almost simultaneously may indicate that the
high-energy photons are produced much further out, downstream along the jet
\citep{valtaoja03,marscher10}. High-resolution observations performed
with the Very Long Baseline Array (VLBA) found that superluminal jet
components are ejected close in time with strong gamma-ray flares
\cite{jorstad01}. \\
With the aim of understanding the origin of gamma-ray emission, 
multiwavelength monitoring campaigns triggered by strong high-energy
flares are required. 
Thanks to the high sensitivity and the gamma-ray all-sky monitoring,
the Large Area Telescope (LAT) on board {\it Fermi} has proved to be a superb
hunter of gamma-ray flares. However, despite the efforts a clear picture
is far from being drawn. From the detailed study of the most variable
sources it seems that not all the flares have the same
characteristics, even if produced within the same source. \\
In this contribution we present results on the multiwavelength
variability study for PKS\,1510-089, 3C\,454.3, and
SBS\,0846+513. These objects underwent several gamma-ray flaring
episodes detected by LAT and were subject of dedicated 
multiwavelength campaigns. In particular these sources are part of the
Monitoring Of Jets in Active galactic nuclei with the VLBA Experiment
(VLBA) programme\footnote{The MOJAVE data archive is maintained at
  http://www.physics.purdue.edu/MOJAVE.}, 
which provides multi-epoch polarimetric 
observations at 15 GHz with a sub-milliarcsecond resolution \cite{lister09}.\\

Throughout this contribution, we assume 
$H_{0} = 71$ km s$^{-1}$ Mpc$^{-1}$, $\Omega_{\rm M} = 0.27$,
$\Omega_{\Lambda} = 0.73$, in a flat Universe.  \\

\section{The flat spectrum quasar PKS\,1510$-$089}

\subsection{Proper motion}

\begin{figure}
\begin{center}
\includegraphics{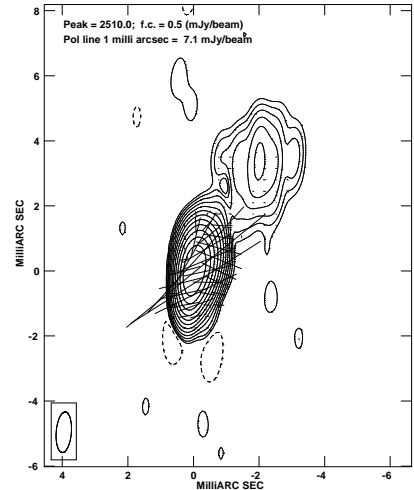}
\vspace{7cm}
\caption{VLBA image at 15 GHz of the parsec-scale structure of
  PKS\,1510$-$089.}
\label{morfo_1510}
\end{center}
\end{figure}

\begin{figure}
\begin{center}
\includegraphics{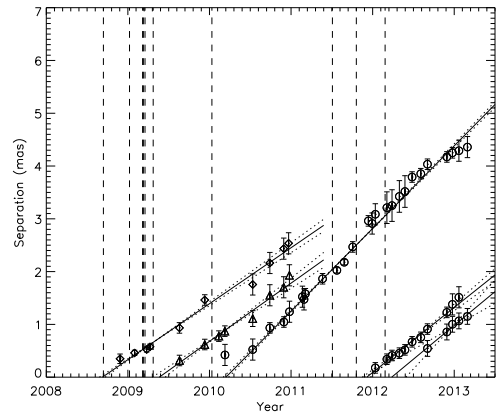}
\vspace{5cm}
\caption{Changes in separation with time between the core component,
  considered stationary, and the new knots ejected since 2008 for
  PKS\,1510$-$089. The 
  solid line represents the regression fit to the 15-GHz VLBA data,
  while the dashed lines represent the uncertainties from the fit
  parameters. Dashed vertical lines indicate the time of the gamma-ray
flares.}
\label{proper_1510}
\end{center}
\end{figure}

PKS\,1510$-$089, at redshift
$z$ = 0.361, is one of the most active blazars observed by {\it
  Fermi}-LAT. Its parsec-scale radio morphology has a core-jet
structure (Fig. \ref{morfo_1510}).
Since 2008 it underwent many gamma-ray flares \citep[see
e.g.][]{mo13,mo11,marscher10,dammando09}. In particular, in October 2011 it
reached an apparent 
isotropic gamma-ray luminosity of 3.7$\times$10$^{48}$ erg/s, 
becoming the second brightest gamma-ray extragalactic object ever
observed by {\it Fermi} \cite{mo13}. Furthermore, 
PKS\,1510$-$089 is one of three flat spectrum radio quasars detected
above 100 GeV \cite{abramowski13,cortina12}.\\
We investigate possible variability in the radio band by means of
multi-epoch MOJAVE data spanning the period 2008-2013. 
The high angular resolution allows us to study changes in
the radio structure, like the emergence of a new superluminal
component close in time with the high-energy flare (see
e.g. \citep{mo13} for the description of the data analysis). \\
During the time interval considered here (2008-2013)
we observed the emergence of 5
superluminal jet components (Fig. \ref{proper_1510}). 
To determine the apparent velocity and
the time of ejection we performed a regression linear fit minimizing
the chi-square error statistics. Results are reported in Table
\ref{1510_tab} and are in good agreement with what was derived in other works
\cite{marscher10, mo11, mo13}. \\ 
The ejection of the new superluminal components occurs close in
time with a gamma-ray flare. However, this correlation does not seem
biunique. A remarkable case is the strong gamma-ray flare
observed in July 2011 which does not seem associated
with the appearance of any new superluminal component, suggesting
possible intrinsic differences in the physics/origin of the different
high-energy flares.\\
         
\begin{table}
\caption{Kinematic properties of the superluminal jet
  components in PKS\,1510$-$089. 
Column 1: component; Col. 2: number of observing epochs
  considered for the linear regression fit; Cols. 3, 4: apparent
  angular velocity (mas/yr) 
and linear velocity, respectively; Col. 5: ejection time
  estimated from the regression fit.}
\begin{center}
\begin{tabular}{ccccc}
\hline
Comp. & Ep.& $\mu$& $\beta$& T$_{0}$\\
\hline
 A& 10& 1.06$\pm$0.06& 23.6$\pm$1.3& 13-09-2008\\
 B&  8& 1.10$\pm$0.11& 24.5$\pm$2.4& 20-05-2009\\
 C& 26& 1.56$\pm$0.03& 34.7$\pm$0.7& 14-03-2010\\
 D& 11& 1.23$\pm$0.11& 27.4$\pm$2.4& 20-11-2011\\
 E&  5& 1.28$\pm$0.40& 28.5$\pm$8.9& 24-03-2012\\
\hline
\end{tabular}
\end{center}
\label{1510_tab}
\end{table}

\subsection{Flux density variability}

The high angular resolution of MOJAVE observations allows us to
separate the emission of the core region from the contribution of
the jet. Disentangling these components is crucial for locating the
region where the high energy emission originates. In
Fig. \ref{1510_multi} we show the 15-GHz light curves for the core
component and the jet knots, together with their fractional polarization
and the electric vector position angle (EVPA). \\
If we consider the
properties of the radio core behaviour, no trivial correlation
between radio and gamma-ray flares is unambiguously determined for all
the episodes (three upper panels in Fig. \ref{1510_multi}). 
Some high-energy flares seem to
precede the increase of the radio flux density, like those observed in
April 2009 and February 2012, as a consequence of opacity effects.
Other flares, like in July 2011, and those observed at the
beginning of 2008, do
not have any obvious relation with changes in the radio emission, 
suggesting that the production of high energy photons
occurs in the innermost part of the AGN, which is totally opaque to
the radio wavelengths. \\
In the case of the October 2011 flare, the high energy outburst seems
to take place almost simultaneously to the millimeter flare. Such
simultaneity may be an indication that high-energy photons are
produced in a region which is already transparent to the radiation at
millimeter wavelength, implying a location that is far away from the
Broad Line Region (BLR), likely parsec scale
downstream along the jet \cite{mo13}. At such distances the high energy
emission would be produced by inverse Compton scattering off the
infrared photons of the dusty torus \cite{sikora08}, or by
synchrotron-self Compton \cite[e.g.][]{marscher10}.\\

\subsection{Polarization properties}

\subsubsection{The unresolved core component}

No clear trend is found between the gamma-ray flares and the
polarization properties. The polarization percentage in the core
never exceeds 5\%, while in the jet knots may reach
10\%. An interesting aspect is shown by the electric vector position
angle (EVPA), both in optical and radio band (Table
\ref{1510_tabpol}). It has been noted that before some gamma-ray
flares, the optical EVPA is subjected to a huge rotation (from
330$^{\circ}$ to 720$^{\circ}$, tracked by daily measurements) which lasts a
few days and culminates when the gamma-ray flare takes place. On the
other hand, the radio EVPA behaves differently during different
flares. For example, although the {\it optical} EVPA rotates of about
380$^{\circ}$ in the July 2011 flaring episode, no significant change
in the {\it radio} EVPA has been observed. A similar behaviour was found in
BL Lacertae and it was interpreted in 
terms of a shock, produced in the very initial part of the jet totally
opaque to the radio wavelengths, which follows a spiral path as it
moves through a toroidal magnetic field \cite{marscher08}.\\
Interestingly, after the February 2012 flare,
which was detected up to very-high-energy (VHE) by MAGIC \cite{cortina12} the
radio EVPA showed a smooth rotation of about 80$^{\circ}$ in 2
months, rather than an abrupt flip, as the one detected after the
other VHE flare that occurred in March and April 2009 and detected by HESS
\cite{abramowski13}, 
when the radio EVPA rotated of
about 70$^{\circ}$ \cite{mo11}.

\subsubsection{The jet knots}

\begin{figure}
\begin{center}
\includegraphics{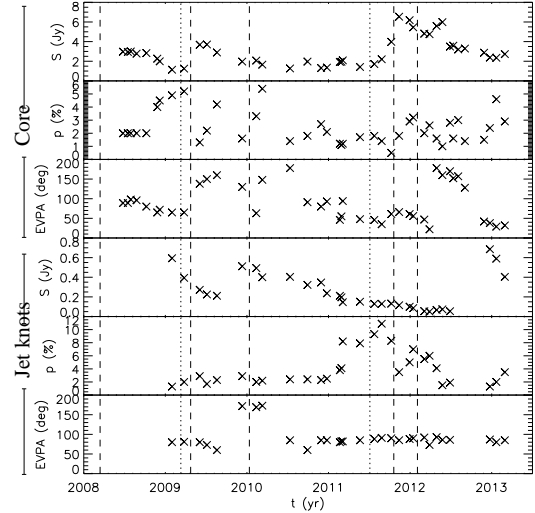}
\vspace{7.5cm}
\caption{Total intensity flux density, fractional polarization, and
  polarization angle for the core component ({\it upper panels}) and
  the ejected knots ({\it lower panels}) in PKS\,1510$-$089. 
Vertical dashed lines and
  dotted lines
  represent gamma-ray flares with or without the emergence of a new jet
  component, respectively.}
\label{1510_multi}
\end{center}
\end{figure}

The high spatial resolution allows us to follow and study the
properties of the various 
ejected knots as they move along the jet. No
correlation is found between the jet properties and the gamma-ray
flares, indicating that the high-energy emission does not arise from
the knot itself, but it is likely located in
the unresolved core component. \\
From Fig. \ref{1510_multi} it is possible to identify three knots 
and follow how
their emission evolves as they propagate along jet.
The flux density decreases as expected for adiabatic losses, and the
polarization percentage increases reaching values up to 10\%, much
larger than what is found in the core component. Remarkably, all the
knots are characterized by the same EVPA $\sim$80$^{\circ}$, with some
exceptions likely due the contamination from another blended component
(Orienti et al., in preparation).\\

\begin{table*}
\caption{Summary of the gamma-ray flares observed in PKS\,1510$-$089. 
Column 1: flare number;
  Cols. 2 and 3: date of the gamma-ray flare and the Astronomer's
  Telegram reporting the detection; Col. 4: change in the radio EVPA,
  from MOJAVE data; Col. 5: change in the optical EVPA with the
  reference: $a$=\cite{marscher10}; $b$=\cite{mo13}; $c$=this
  contribution (Orienti et al., in prep.); Col. 6: ejection of a
  superluminal knot close in time; Col. 7: VHE detection and
  reference: $d$=\cite{abramowski13}; $e$=\cite{cortina12}. }
\begin{center}
\begin{tabular}{ccccccc}
\hline
Flare & Date& ATel & $\Delta$ EVPA& $\Delta$ EVPA & knot & VHE \\
      &     & N.   &   radio      &  optical      &      &     \\ 
\hline
1 & 13/09/2008 &\#1743& $<$15$^{\circ}$&    -         & Yes  & -  \\
2 & 09/03/2009 &\#1957&  $<$10$^{\circ}$&    -         & No   &Yes, d  \\
3 & 25/04/2009 &\#2033&   $\sim$70$^{\circ}$&  720$^{\circ}$, a& Yes &
Yes, d \\
4 & 11/01/2010 &\#2385& $<$10$^{\circ}$&   -           & Yes  & -  \\
5 & 04/07/2011 &\#3473& $<$10$^{\circ}$&  380$^{\circ}$, b & No   & -  \\
6 & 19/10/2011 &\#3694& $<$15$^{\circ}$&  -           & Yes  & -  \\
7 & 28/01/2012 &\#3907& $\sim$80$^{\circ}$& 330$^{\circ}$, c& Yes & Yes, e \\
\hline
\end{tabular}
\end{center}
\label{1510_tabpol}
\end{table*}

\begin{figure}
\begin{center}
\includegraphics{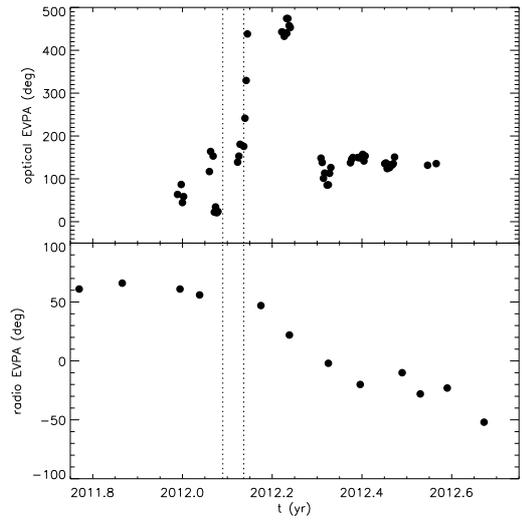}
\vspace{6.5cm}
\caption{Optical EVPA ({\it top}) and radio EVPA ({\it
    bottom}) in PKS\,1510$-$089. 
Multiples of 180$^{\circ}$ are added to EVPA as needed to
  minimize the jumps in consecutive values. 
The area between the vertical lines represents the period of the VHE
detection by MAGIC \cite{cortina12}. Optical data are from the Steward
Observatory blazar monitoring programme of the University of Arizona.
A description of this monitoring project, the calibration and the data products can be found in \cite{smith09}. }
\label{1510_evpa}
\end{center}
\end{figure}

\begin{figure}
\begin{center}
\includegraphics{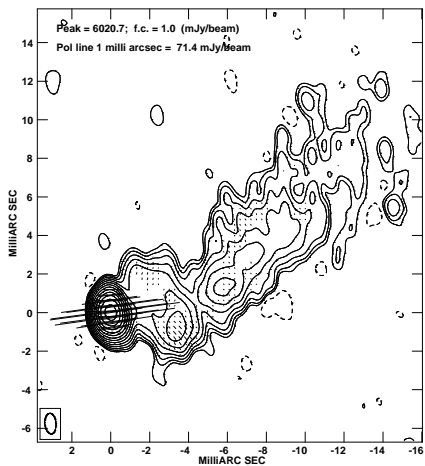}
\vspace{7cm}
\caption{VLBA image at 15 GHz of the parsec-scale structure of
  3C\,454.3.}
\label{morfo_3c454}
\end{center}
\end{figure}

\section{The flat spectrum quasar 3C\,454.3}

This source, at redshift $z$ = 0.859,  
was the most active blazar in
gamma rays during the first three 
years of {\it Fermi} observations. On
the parsec scale its
radio emission has a core-jet structure (Fig. \ref{morfo_3c454}). 
Since 2008, it has undergone three main gamma-ray
outbursts \cite{ackermann10,abdo11}. 
In particular, in November 2010 it underwent an
extraordinary 5-day gamma-ray flare and became the brightest object in
the gamma ray sky, 5 times brighter than the Vela pulsar
\cite{abdo11}. After this major outburst the source entered in a quiet
activity state at high energy, providing us with a unique opportunity for
investigating a correlated variability between radio and gamma
rays. \\

\subsection{The flux density variability}

\begin{figure}
\begin{center}
\includegraphics{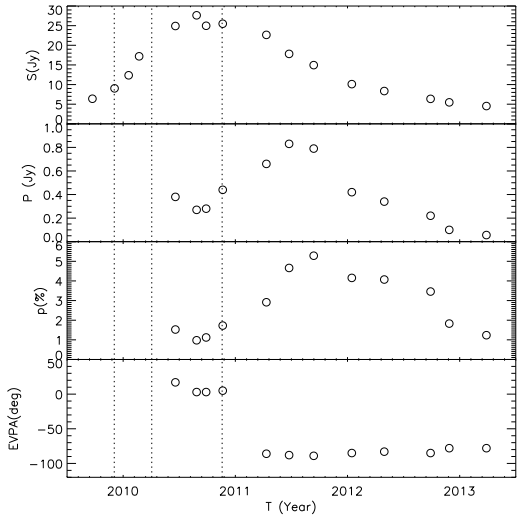}
\vspace{7cm}
\caption{From top to bottom: 
total intensity flux density, polarization flux density, 
fractional polarization, and
  polarization angle for 3C\,454.3, as derived from 15-GHz MOJAVE
  data. Vertical lines indicate the time of the gamma-ray flares.}
\label{3c454_totpol}
\end{center}
\end{figure}

\begin{figure}
\begin{center}
\includegraphics{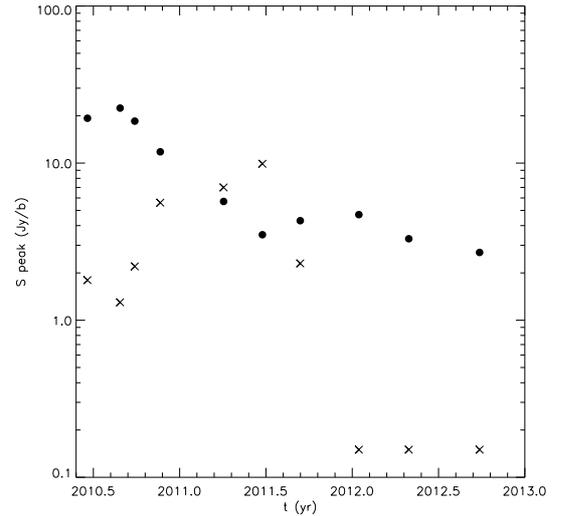}
\vspace{7cm}
\caption{Peak flux density of the core component ({\it filled
    circles}), and of the jet knot ({\it crosses}) for 3C\,454.3.}
\label{3c454_peak}
\end{center}
\end{figure}

To investigate the radio properties of the central region of
3C\,454.3, we analysed MOJAVE data obtained between June 2009 and July 2013.\\
The analysis of the radio emission shows that the flux density was in
a local minimum (S$_{\rm 15 GHz} \sim 6.4$ Jy) 
before the first gamma-ray flare detected in December 2009,
and then it started to increase and reached its maximum (S$_{\rm 15
  GHz} \sim$ 27.6 Jy on mas-scale) in August 2010, about four months after a 
second strong gamma-ray flare that took place in April
2010 (Fig. \ref{3c454_totpol}). 
Unfortunately the time sampling of the observations is not
adequate to investigate the flux density behaviour after the huge
gamma-ray flare in November 2010. However, VLBI observations at higher
frequencies pointed out an additional rising of the radio flux density
after the November 2010 flare with the 43 GHz increase leading on the
22 GHz, as expected in presence of opacity effects
\cite[e.g.][]{wehrle12,nagai13}.\\
The analysis of multi-epoch 43-GHz VLBA observations could reveal the
ejection of two superluminal jet components: one close in time with the
December 2009 flare, and the other likely associated with the November
2010 \cite{jorstad13}. No new superluminal component was observed
associated with 
the April 2010 flare.\\ 
Since 2011 the source has been in a quiet activity state in gamma rays
without any further flare episode. The same behaviour is observed in the 15-GHz
light curve, where the flux density has decreased to about 4 Jy. This
result suggests that the variability at the two edges of the
electromagnetic spectrum is tightly related, at least in this
source. A different behaviour is observed, for example, in the
radio-loud Narrow Line Seyfert 1 SBS\,0846+513 where outbursts in the
radio band are observed during quiet activity state in gamma rays (see
Section 4) \cite{dammando12}.\\
Thanks to the high resolution of the VLBA data, we could separate the
core emission from that arising from the compact jet component that
was ejected at the end of 2009, close in time with the gamma-ray flare
detected in December 2009. The analysis of the light curves of the
core and the jet component shows that the flux density increase may be
assigned to the knot component, which becomes even brighter than the
unresolved core region (Fig. \ref{3c454_peak}). 
This result is in agreement with the 43-GHz
VLBA data presented in \cite{jorstad13}.\\

\subsection{The polarized emission}

An interesting behaviour is shown by the polarized emission. 
The polarized flux
density reached its maximum when the total emission is already
decreasing. Furthermore, after the November 2010 flare, the radio EVPA
had an abrupt flip of almost 90$^{\circ}$, switching from
$\sim$5$^{\circ}$ to $\sim$95$^{\circ}$. This rotation may be
due to the transition between the optically thick and thin
regime. This interpretation is supported by observations at 43 GHz,
where the 90$^{\circ}$ flip of the EVPA was already observed in
November 2009, when at 15 GHz the EVPA was still $\sim$5$^{\circ}$
\cite{wehrle12} (Faraday rotation effects are negligible at these
frequencies). From the analysis of the EVPA we find that the
magnetic field is perpendicular to the jet axis. This is in agreement
with the presence of a shock that is propagating along the jet and
is ordering the component of the magnetic field which is perpendicular
to the shock propagation. This may also explain the increase of the
polarized emission, since the degree of polarization is related to the
ordered component of the magnetic field which is enhanced by the
passage of the shock.\\

\section{The NLSy1 SBS\,0846+513}

\begin{figure}
\begin{center}
\includegraphics{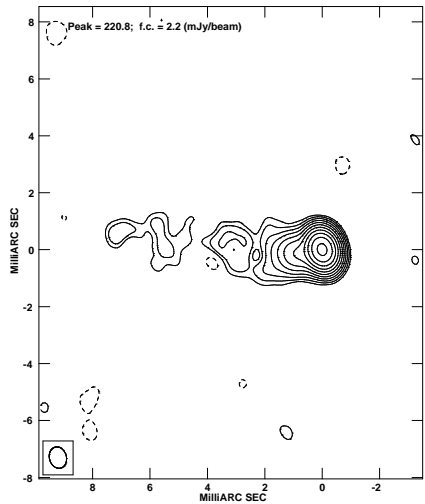}
\vspace{7cm}
\caption{VLBA image at 15 GHz of the NLSy1 SBS\,0846+513.}
\label{morfo_0846}
\end{center}
\end{figure}

The radio source SBS\,0846+513 is associated with a Narrow Line
Seyfert 1 galaxy at redshift $z=0.5835$. This is one of the five
NLSy1s detected by {\it Fermi}-LAT. Its radio emission has a core-jet
structure (Fig. \ref{morfo_0846}).
The source underwent three strong
gamma-ray flares: one in June 2011, when
the source was detected for the first time at high energies
\cite{dammando12}, and the other two in May and August 2012
\cite{dammando13}. \\
The detection of gamma rays from this class of sources was unexpected,
opening new questions on the physics in these objects \cite{dammando13a}. \\
To investigate the nature of this object we compare the gamma-ray
light curve with the data at 15 GHz from the Owens Valley Radio
Observatory (OVRO). From this comparison 
we could not find any clear correlation between the
variability in the two energy bands. Gamma-ray flares are not
reliably associated with radio outbursts: only a slight increase of the flux
density at 15 GHz is observed a few months after the gamma ray
flare. On the other hand, a large radio outburst occurred in December 2009,
when the source was not detected in gamma rays. Interestingly, the
analysis of the multi-epoch VLBA data from the MOJAVE programme
pointed out that a
superluminal jet component was ejected close in time with the radio
outburst, representing an example of knot emission not connected to any
gamma ray flaring activity \cite{dammando13}.\\
A complicated behaviour is observed in 2012. Three radio outbursts
occurring in May, October, and December 2012 were
detected at 15 GHz by OVRO, while only two gamma-ray flares, one in
May and the other in August 2012, were observed. The most likely
scenario suggests that the radio outbursts detected in October and
December 2012 are likely the delayed counterparts of the two gamma-ray
flaring episodes, while the radio flare observed in May has no clear
counterpart at high energy \cite{dammando13}.

\section{Concluding remarks}

Multiwavelength monitoring campaigns suggest a relation between gamma-ray
flares and the radio variability, explained in terms of a shock moving
along the jet, whose manifestation is a superluminal knot observable
with high-frequency VLBI observations. However, the variability
observed in gamma rays does not always show the same properties in the
other energy bands, even if we consider the same source.\\
So far only a handful of the most powerful objects have been studied,
and no statistically complete conclusions can be drawn. To better
understand the physics at work in these objects, larger samples of
objects spanning a wider luminosity range is needed. \\
With the jump in
sensitivity provided by ALMA (more than an order of magnitude with
respect to current interferometers working at same wavelengths) we
will be able to study a large number of faint objects, both in total
intensity and polarization, which is fundamental for understanding the
physics of high energy emitting radio sources.\\

\begin{acknowledgement}
The Fermi-LAT Collaboration acknowledges generous ongoing support from
a number of agencies and institutes that have supported both the
development and the operation of the LAT as well as scientific data
analysis. These include the National Aeronautics and Space
Administration and the Department of Energy in the United States, the
Commissariat à l'Energie Atomique and the Centre National de la
Recherche Scientifique / Institut National de Physique Nucléaire et de
Physique des Particules in France, the Agenzia Spaziale Italiana and
the Istituto Nazionale di Fisica Nucleare in Italy, the Ministry of
Education, Culture, Sports, Science and Technology (MEXT), High Energy
Accelerator Research Organization (KEK) and Japan Aerospace
Exploration Agency (JAXA) in Japan, and the K. A. Wallenberg
Foundation, the Swedish Research Council and the Swedish National
Space Board in Sweden. Additional support for science analysis during
the operations phase is gratefully acknowledged from the Istituto
Nazionale di Astrofisica in Italy and the Centre National d’Études
Spatiales in France. This research has made use of the data from the MOJAVE data base that is maintained by the MOJAVE team (Lister et al. 2009b).
Data from the Steward Observatory spectropolarimetric project were used. This programme is supported by Fermi Guest Investigator grants NNX08AW56G and NNX09AU10G.  \\

\end{acknowledgement}

\end{document}